\documentclass[preprint,12pt]{elsarticle}

\usepackage{amssymb}
\usepackage{amsmath}

\journal{EANM Innovation}

\usepackage{graphicx}%
\usepackage{multirow}%
\usepackage{amsmath,amssymb,amsfonts}%
\usepackage{amsthm}%
\usepackage{mathrsfs}%
\usepackage[title]{appendix}%
\usepackage{xcolor}%
\usepackage{textcomp}%
\usepackage{manyfoot}%
\usepackage{booktabs}%
\usepackage{algorithm}%
\usepackage{algorithmicx}%
\usepackage{algpseudocode}%
\usepackage{listings}%

\usepackage{setspace}

\usepackage{isotope}
\usepackage{graphicx,tabularx,booktabs,colortbl}
\usepackage{pgfplotstable}
\usepackage{subcaption}
\usepackage{natbib}
\usepackage[load-configurations=abbreviations, separate-uncertainty=true, multi-part-units=single]{siunitx}
\usepackage[T1]{fontenc}
\usepackage[colorlinks=false]{hyperref}

\newcommand{\eq}{\; = \;}

\newcommand{\tb}{\isotope[149]{Tb}}
\newcommand{\gd}{\isotope[149]{Gd}}
\newcommand{\revised}[1]{\textcolor{black}{#1}}
\newcommand{\rrevised}[1]{\textcolor{black}{#1}}

\newcommand{\rmax}{RC_{max}}
\newcommand{\rmean}{RC_{mean}}
\newcommand{\rpeak}{RC_{peak}}
\newcommand{\pg}{p$\gamma$}
\newcommand{\cov}{\mathrm{COV} }

\DeclareSIUnit\year{y}

\begin{document}

\begin{frontmatter}

\title{Terbium-149 PET/CT: First Quantitative Imaging with a Clinical Long Axial Field-of-View Scanner}

\author[1,2,3]{Lorenzo Mercolli}
\author[4]{Pascal V.\ Grundler}
\author[4]{Anzhelika N.\ Moiseeva}
\author[3]{Lars Eggimann}
\author[3]{Saverio Braccini}
\author[5]{\revised{Ulli Köster}}
\author[4,6]{Nicholas P.\ van der Meulen}

\affiliation[1]{organization={Department of Nuclear Medicine, Inselspital, Bern University Hostpital, University of Bern},
            city={Bern}, country={Switzerland}}
\affiliation[2]{organization={ARTORG Center for Biomedical Engineering Research, University of Bern},
            city={Bern}, country={Switzerland}}
\affiliation[3]{organization={Albert Einstein Center for Fundamental Physics (AEC), Laboratory for High Energy Physics (LHEP), University of Bern},
            city={Bern}, country={Switzerland}}
\affiliation[4]{organization={Center for Radiopharmaceutical Sciences, PSI Center for Life Sciences},
            city={Villigen-PSI}, country={Switzerland}}
\affiliation[6]{organization={Institut Laue-Langevin}, 
            city={Grenoble}, country={France}}
\affiliation[6]{organization={Laboratory of Radiochemistry, PSI Center for Nuclear Engineering and Sciences}, 
            city={Villigen-PSI}, country={Switzerland}}

\begin{abstract}

\textbf{Introduction:}
Terbium-149 (\tb) is a promising radionuclide for targeted $\alpha$ therapy that has a non-zero branching ratio (BR) for positron decay. However, its relatively low positron branching fraction and multiple prompt $\gamma$ emissions may challenge quantitative imaging. This study evaluates the imaging performance and quantitative accuracy of \tb\ using a clinical long axial field-of-view (LAFOV) PET/CT system.

\textbf{Methods:}
Quantitative accuracy of \tb\ was assessed with a NEMA IEC body phantom, which was filled with about \SI{45}{\mega\becquerel} \tb\ and a sphere-to-background ration of 10:1. The phantom was scanned for \SI{20}{\minute} and shorter scan times and lower activities were simulated. Recovery coefficients, coefficient of variation, and lung residual error were evaluated for different reconstruction settings and compared to the EARL standard 2 for \isotope[18]{F}.

\textbf{Results:}
High-quality PET images of \tb\ were obtained, even with a simulated total activity of \SI{4.5}{\mega\becquerel}. The \SI{20}{\minute} and full activity scan yielded a mean recovery coefficient $RC_\textit{mean}$ of $0.55$, $0.69$, $0.73$, $0.76$, $0.79$, and $0.81$ for the six phantom spheres. Despite the low count statistics, the coefficient of variation stays mostly below $15\,\%$. Relative scatter correction combined with prompt $\gamma$ modeling provided robust quantification.  

\textbf{Conclusion:}
\tb\ can be imaged using a commercial LAFOV PET/CT with a quantitative accuracy comparable to the EARL standard 2 for \isotope[18]{F}. These findings demonstrate the feasibility of PET-based treatment verification and dosimetry for targeted $\alpha$ therapy with \tb.
\end{abstract}


\begin{highlights}
\item First imaging of \tb\ with a commercial PET/CT scanner.
\item PET imaging of an $\alpha$ emitting therapeutic radionuclide. 
\item Quantitative accuracy of \tb\ comparable to \isotope[18]{F}. 
\end{highlights}

\begin{keyword}
Terbium-149 \sep PET imaging \sep Long axial field-of-view PET/CT \sep Quantification \sep Targeted $\alpha$ therapy
\end{keyword}

\end{frontmatter}

\newpage 

\section*{List of abbreviations}

\begin{table}[h!]
    \begin{tabular}{ll}
        AC & Attenuation correction \\
        BR & Branching ratio \\
        COV & Coefficient of variation \\
        CT & Computed tomography \\
        FWHM & Full width at half maximum \\
        FWTM & Full width at tenth of maximum \\
        HPGe & High-purity germanium spectrometer \\ 
        LAFOV & Long axial field-of-view \\
        PET & Positron emission tomography \\
        \pg\ & Prompt photons \\
        PR & Positron range \\
        PSF & Point spread function \\
        ROI & Region-of-interest \\
        SAFOV & Standard axial field-of-view \\
        SPECT & Single-photon emission computed tomography \\
        TOF & Time-of-flight \\
        VOI & Volume-of-interest \\
    \end{tabular}
\end{table}

\section{Introduction}

\isotope[68]{Ga}- and \isotope[177]{Lu}–labeled prostate-specific membrane antigen ligands and DOTA-conjugated peptides have achieved a landmark success as theranostic agents and have become a cornerstone modern nuclear medicine. However, chelation is different for these diagnostic and therapeutic radionuclides and the post-treatment imaging of \isotope[177]{Lu} relies on single-photon emission computed tomography (SPECT). Terbium has emerged as a promising rare earth with isotopes that are suitable for both diagnostic and therapeutic applications \cite{Moiseeva2024,mueller2024}. Some of the terbium isotopes have already found their way into clinics \cite{Zhang2019,Baum2021,Fricke2025}.

The combination of $\beta^-$ emission with conversion and Auger electrons of \isotope[161]{Tb} increases linear energy transfer (LET) compared to \isotope[177]{Lu}, to the patients' benefit. The use of \tb\ could further extend this approach by adding $\alpha$-particles to the treatment possibilities \cite{beyer2004,mueller2014,mueller2016,favaretto2024,Mapanao2025,Tosato2025}. However, $\alpha$ emitting radionuclides, like \isotope[223]{Ra} or \isotope[225]{Ac}, are notoriously difficult to image \cite{Owaki2017,Seo2019,Tulik2024} and rely on SPECT. SPECT is in fact the standard imaging modality for post-treatment verification and image-based dosimetry, but it suffers from limitations in sensitivity, spatial resolution and quantitative accuracy. In addition to the $\alpha$ decay channel, \tb\ has a positron branching ratio (BR) of $7.12\pm0.24 \%$ (see Tab.~\ref{tab:isotope}). \tb\ therefore offers the ability to perform post‑therapy control and dosimetry with positron emission tomography (PET), which represents a potential game changer in nuclear medicine. In comparison with \isotope[90]{Y}, the only therapeutic radionuclide imaged with PET in clinical routine, \tb\ imaging is not hampered by a very low positron BR \cite{Elschot2013,Zeimpekis2023,Mercolli2024}. 
Unlike SPECT, PET provides superior sensitivity, resolution, and quantitative accuracy. This would allow clinicians to assess on- and off‑target uptake and estimate patient‑specific radiation doses with unprecedented reliability. By integrating PET into a post-therapy workflow, post‑therapy imaging would evolve from a qualitative check into a powerful quantitative tool.

In this study, we image \tb\ for the first time using a clinical long axial field-of-view (LAFOV) PET/CT system and we investigate the quantitative accuracy. We compare the pertinent recovery coefficients to clinical standard for \isotope[18]{F}. The decay properties of \tb\ are summarized in Tab.~\ref{tab:isotope}. The low positron BR of \tb\ in comparison with \isotope[18]{F} or \isotope[68]{Ga} motivates the use of a LAFOV PET/CT \cite{Mehadji2025,Alberts2021,Prenosil2022,Spencer2021,Badawi2019,Karp2020}. \revised{The increased sensitivity of LAFOV PET/CT compared to standard axial field-of-view (SAFOV) systems possibly compensates for low positron BR and low activities or shorten scan times.} The expected value of the positron's kinetic energy is much higher for \tb\ than for \isotope[18]{F}. We therefore test the vendor's positron range (PR), or rather point spread function (PSF), correction method as well as the prompt photon (\pg) correction algorithm, given the large number of $\gamma$ lines reported in Tab.~\ref{tab:isotope} (see e.g.\ \cite{Conti2016} for a discussion on \pg\ in PET imaging).

\begin{table}[htb!]
    \centering
    \caption{Decay properties of \tb\ in comparison to other PET radionuclides used in clinical practice, as retrieved from \url{www-nds.iaea.org} (only \pg\ lines with a BR above than $1\%$ are reported).}
    \label{tab:isotope}
    {\footnotesize
    \begin{tabular}{lcccc}
        \toprule
        Nuclide & Half-life $[\SI{}{\min}]$ & $BR_{\beta^+}$ $[\%]$ & $E_\gamma \; [\SI{}{\kilo\electronvolt}]$ & $BR_\gamma$ $[\%]$ \\
        \midrule
        \rowcolor{lightgray} \isotope[18]{F} & $109.77\pm0.05$ & $96.73\pm0.04$ & - & - \\
        \isotope[68]{Ga} & $67.71 \pm 0.08$ & $88.91 \pm 0.09$ & $1077.34\pm 0.05$ & $3.22\pm 0.03$  \\
        \rowcolor{lightgray} \isotope[82]{Rb} & $1.2575 \pm 0.0002$ & $95.27 \pm 0.57$ & $776.511\pm0.010$ & $15.1\pm0.3$ \\ 
        & & & $602.73\pm 0.08$ & $62.9\pm 0.7$ \\
        & & & $722.78\pm 0.08$ & $10.36 \pm 0.12$ \\
        & & & $ 1325.52\pm 0.08$ & $1.578 \pm 0.021$ \\
        & & & $1376.09\pm0.08$ & $1.79\pm 0.03$ \\
        & & & $1509.36\pm0.08$ & $3.25\pm 0.05$ \\
        \multirow{-6}{*}{\isotope[124]{I} } & \multirow{-6}{*}{$6013.44 \pm 0.432$ } & \multirow{-6}{*}{$22.69 \pm 1.33$ } & $1690.96\pm0.08$ & $11.15\pm0.17$ \\
        \rowcolor{lightgray} & & & $164.98 \pm0.02$ & $26.7 \pm0.6$ \\
        \rowcolor{lightgray} & & & $187.22\pm0.02 $ & $4.36\pm0.11$ \\
        \rowcolor{lightgray} & & & $352.24\pm0.02$ & $29.8\pm0.7$ \\
        \rowcolor{lightgray} & & & $388.57\pm0.02$ & $18.6\pm 5$ \\
        \rowcolor{lightgray} & & & $464.85\pm0.02$ & $5.73 \pm 0.15$ \\
        \rowcolor{lightgray} & & & $652.12 \pm0.02$ & $16.5\pm0.4$ \\
        \rowcolor{lightgray} & & & $772.65\pm0.03 $ & $1.63 \pm0.05$ \\
        \rowcolor{lightgray} & & & $817.1 \pm 0.2$ & $11.8 \pm3$ \\
        \rowcolor{lightgray} & & & $853.43\pm0.01 $ & $15.7\pm 0.4$ \\
        \rowcolor{lightgray} & & & $861.86\pm 0.02 $ & $7.60\pm0.19$ \\
        \rowcolor{lightgray} & & & $1040.65 \pm0.04$ & $1.48\pm0.06$ \\
        \rowcolor{lightgray} & & & $1135.3\pm0.1$ & $1.20 \pm 0.05$ \\
        \rowcolor{lightgray} & & & $1175.4 $ & $	3.31 \pm  15$ \\
        \rowcolor{lightgray} & & & $1341.19 \pm 0.06$ & $2.33\pm 0.10$ \\
        \rowcolor{lightgray} & & & $1640.26\pm0.06$ & $3.22\pm 11$ \\
        \rowcolor{lightgray} \multirow{-16}{*}{\isotope[149]{Tb}} & \multirow{-16}{*}{$247.2 \pm 1.8$} & \multirow{-16}{*}{$7.12\pm0.24$ } & $1827.5 $ & $	1.11\pm 0.06$ \\
        \bottomrule
    \end{tabular}
    }
\end{table}




\section{Materials and methods}

A tantalum target was irradiated with a \SI{1.4}{\giga\electronvolt} proton beam and produced radionuclides were mass separated at the ISOLDE facility (project IS688) of CERN (Geneva, Switzerland). Mass 149 isobars and molecular pseudo-isobars were implanted into zinc-coated foils and the desired radionuclide chemically separated at the Paul Scherrer Institute (PSI, Villigen-PSI, Switzerland) to obtain a high-purity product ($>99.7\%$ radionuclidic purity \SI{41}{\min} after the end of separation), as described in Ref.~\cite{favaretto2024}. A glass vial with about \SI{186.3}{\mega\becquerel} \tb\ was sent to the Department of Nuclear Medicine at Inselspital (Bern, Switzerland) where the data acquisition was performed with a Biograph Vision Quadra (Siemens Healthineers, Knoxville TN, USA), henceforth referred to as Quadra. There is no preset for acquiring \tb\ data on Quadra. Hence, the scans were performed in the \isotope[44]{Sc} setting \cite{Lima2020}, which requires a correction for the positron fraction as well as for the decay factor in the post-processing of the images. All image reconstructions were performed off-line using the prototype software e7-tools (Siemens Healthineers, Knoxville TN, USA). 

The dose calibrator in the Department of Nuclear Medicine (VDC-405/VIK-202, Comecer, Italy) was cross-calibrated with the one at PSI. However, the rather complex decay chain of \tb\ and the out-of-equilibrium activity distribution of the daughter radionuclides make an activity measurement of \tb\ with a dose calibrator challenging. We therefore opted to measure \tb\ activities by sampling the phantom compartments and measure the samples with a high-purity germanium gamma spectrometer (HPGe). \revised{With the known detector efficiency, the activities in the samples were determined through a background-subtracted fit to the relevant photopeaks. } 

\revised{We prepared capillaries (outer diameter \SI{1}{\mm} and inner diameter \SI{0.6}{\mm} with few \SI{}{\kilo\becquerel} of \tb) and a NEMA IEC body phantom (Data Spectrum Corporation) for the assessment of the imaging capabilities of Quadra with \tb}. Following the NEMA NU 2-2018 protocol, the capillaries were measured at the center and at one eighth of the axial field-of-view (FOV) and at \SIlist{1;10;20}{\cm} in the transverse plane. The images of the capillaries were reconstructed with a filtered backprojection algorithm and $880\times880$ matrix size. \revised{Reconstructing capillaries with an iterative reconstruction algorithm would carry the risk of strong non-linearity and possibly Gibbs-type artifacts}. The reconstruction includes time-of-flight (TOF) but no attenuation (AC), scatter nor point-spread-function (PSF) corrections were applied. The spatial resolution of Quadra with \tb\ is reported according to the NEMA NU 2-2018 standard in terms of full width at half maximum (FWHM) and full width at tenth maximum (FWTM). \revised{Both metrics were determined through an interpolation of the line profiles in the corresponding direction.}

\begin{table}[htb!]
    \caption{Details of the NEMA phantom preparation at scan time.}\label{tab:activities}
    \centering
    {\footnotesize
    \begin{tabular}{llcc} 
    \toprule 
    & & HPGe [\SI{}{\kilo\becquerel\per\ml}] &  Ratio to \tb \\ 
    \midrule 
    \rowcolor{lightgray} & \tb & $33.7\pm0.9$ &  1  \\ 
    \rowcolor{lightgray}  Spheres & \isotope[149]{Gd} & $0.98\pm0.06$ &  $0.0290\pm0.0019$ \\ 
    \rowcolor{lightgray}  & \isotope[145]{Eu} & $0.360\pm0.027$ &  $0.0107\pm0.0009$ \\ 
    \multirow{3}*{Background} & \tb & $2.85\pm0.08$ &  1 \\ 
    & \isotope[149]{Gd} & $0.10\pm0.01$  & $0.035\pm0.004$ \\ 
    & \isotope[145]{Eu} & $0.027\pm0.004$ & $0.0093\pm0.0015$ \\ 
    \bottomrule 
    \end{tabular}
    }
\end{table}

\begin{figure}
    \centering
    \includegraphics[width=\linewidth]{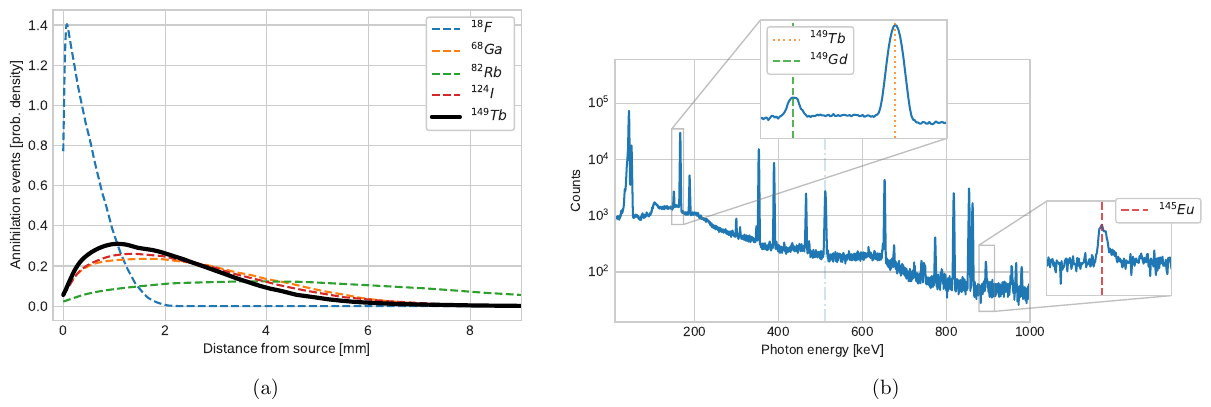}     
    \caption{\revised{(a) probability of positron annihilation in water for point sources of different radionuclides. (b) HPGe spectrum of the stock solution sample with the details of the \tb\ peak at \SI{164.98 \pm 0.02}{\kilo\electronvolt} and \gd\ peak at \SI{149.730 \pm 0.010}{\kilo\electronvolt} as well as for \isotope[145]{Eu} at \SI{893.73 \pm 0.03}{\kilo\electronvolt}. }}
    \label{fig:hpge_spectrum}
\end{figure}

\revised{The NEMA phantom was filled targeting a sphere-to-background activity concentration ratio of about $10:1$, in line with the EARL standard 2 for \isotope[18]{F} \cite{Boellaard2008,Boellaard2010,Kaalep2018,Kaalep2019}.} The total \tb\ activity in the NEMA phantom at scan time was \SI{45.29\pm0.10}{\mega\becquerel}. The HPGe measurements of a $\SI{1}{\ml}$ aliquot from the stock solution, i.e.\ phantom spheres, and from the background compartment yielded activity concentrations reported in Tab.~\ref{tab:activities}. The top right panel of Fig.~\ref{fig:hpge_spectrum} shows the spectrum measured, with a detailed view of the peaks that were used to determine the activities in the aliquot. A HPGe measurement at PSI \SI{41}{\min} after the chemical separation showed a \gd/\tb\ activity ratio of about $0.0022$, which is consistent with Tab.~\ref{tab:activities} at scan time (about \SI{5.42}{\hour} after the end of chemical separation). Despite having \gd\ and \isotope[145]{Eu} activities at the percent level of the \tb\ activity at scan time, we neglected the daughter's contribution to the positron BR. \gd\ has a BR of $<0.0033 \%$ while \isotope[145]{Eu} has a BR of $1.91\pm0.05 \%$. This means that the contribution of \gd\ and \isotope[145]{Eu} to the positron emission is less than the error on the positron BR of \tb, which is about $3.3\%$ (see Tab.~\ref{tab:isotope}).

\begin{figure} 
\centering
    \includegraphics[width=\linewidth]{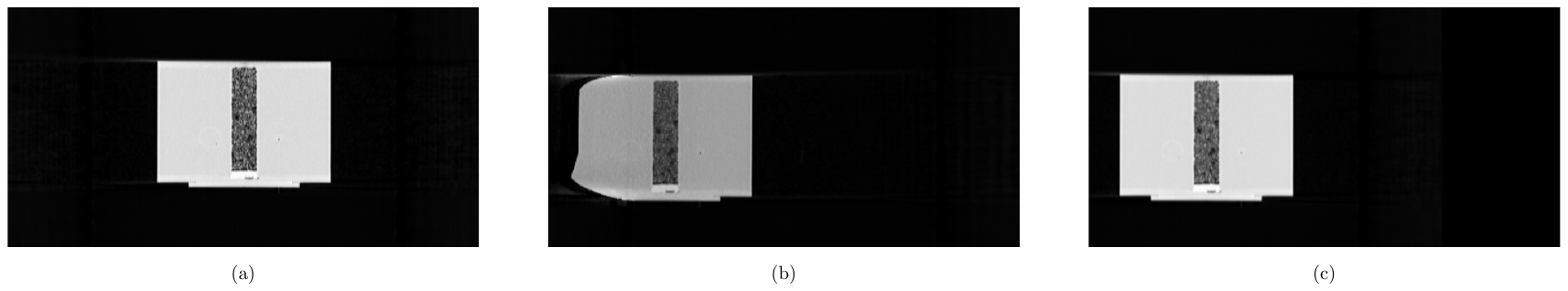}
    \caption{\revised{(a) CT image of the NEMA phantom centered in the axial direction. (b) CT image with the off-axis NEMA phantom with visible spatial distortions. (c) undistorted off-axis CT image of the NEMA phantom.}}
    \label{fig:ac_ct}
\end{figure}

\revised{Reconstructions of the NEMA phantom data were performed with the vendors proprietary ordered subset expectation maximization (OSEM) algorithm including TOF, attenuation and scatter correction, as well as PSF recovery. The randoms sinogram is estimated through a delayed time window.} The matrix size was $440\times 440$ and a \SI{2}{\mm} Gaussian FWHM post-reconstruction filter was applied. We compared different scan times, number of iterations and scatter correction models (absolute vs.\ relative \cite{Bal2021}). \revised{Four subsets were used for all reconstructions}. In the case of absolute scatter correction, no \pg\ correction is computed. Since the activity of $\alpha$ emitters, like \isotope[225]{Ra} or \isotope[225]{Ac}, that are administered to patients is typically limited to a few \SI{}{\mega\becquerel} (though these radionuclides have a significantly longer half-life compared to \tb), we simulated a \SI{20}{\min}-scan with a low activity by downsampling the list mode data to one tenth of the acquired raw data. This brings the total activity in the NEMA phantom to about \SI{4.5}{\mega\becquerel}, which is in the order of administered activities of $\alpha$ emitters used in clinical routine. The resulting reconstructions are marked as `\SI{20}{\min} trim'.

The positron's kinetic energy from the \tb\ decay is significantly higher than for \isotope[18]{F}, which leads to an increased PSF and should therefore be corrected.  Using the FLUKA particle transport toolkit \cite{fluka1,fluka2,fluka3,flair} with a modified MGDRAW routine \rrevised{(PRECISIOn in DEFAULTS card)}, we simulated the location of positron annihilation from a point source of \tb\ in water. The resulting kernel density is used to correct for the PR in the NEMA phantom images. \revised{Panel (a) of Fig.~\ref{fig:hpge_spectrum} shows the PR in water of \tb\ in comparison with other radionuclides used in clinical routine. }
Instead of using custom PR correction projectors for the OSEM algorithm, as was done e.g.\ in Ref.~\cite{Kertesz2022} for \isotope[124]{I}, we applied a Richardson–Lucy deconvolution to the reconstructed images. 
This allowed us to correct the PSF for the PR without compromising the scanner dependent contribution to the PSF \cite{Panin2006} and without the need of a custom implementation of the OSEM reconstruction algorithm. As is common in clinical reconstruction protocols, we applied \SI{2}{\mm} Gauss filter after the deconvolution with the simulated PR kernel. 

As reported in Tab.~\ref{tab:isotope} and illustrated in Fig.~\ref{fig:hpge_spectrum}, \tb\ emits a large number of \pg\ (see e.g.\ cite{Conti2016} for a review on imaging non-pure positron emitting-radionuclides). e7tools offers a \pg\ correction algorithm, which fits a linear combination of the smoothed randoms sinogram and the scatter contribution in the tail region of the emission sinogram \cite{Conti2016,Bal2021}. This \pg\ correction depends on Quadra's relative scatter correction algorithm described in Ref.~\cite{Bal2021}. For absolute scatter correction no \pg\ correction is applied. To challenge Quadra's \pg\ correction method, we scanned the NEMA phantom with a large transverse displacement. In this position, we expected the \pg\ correction model to fail, given the reduced distance to the detector ring. We noticed that the CT was distorted at the border of the FOV, as can be seen in \revised{panel (b) of Fig.~\ref{fig:ac_ct}}. In order to disentangle the limitations of the \pg\ correction and the CT distortions, we constructed a synthetic CT by registering the undistorted phantom from the central scan to the off-axis image (see \revised{panel (c) Fig.~\ref{fig:ac_ct}}). Reconstruction with this synthetic AC CT are labeled as `mod. CT'. 

We evaluated the quantitative accuracy of \tb\ on Quadra by following the EARL standard 2 \citep{Boellaard2008,Boellaard2010,Kaalep2018,Kaalep2019}. The EARL compliance of Quadra for \isotope[18]{F} has been shown in Ref.~\cite{Prenosil2022a}. In this study, we compare the three recovery coefficients $\rmax$, $\rmean$ and $\rpeak$ with the EARL standard 2 limits for \isotope[18]{F}. On the PET image, the maximum pixel values $A_{max}^{(i)}$ in each sphere $s_i$ $ i=1,...,6$ of the phantom is identified and therefrom a background corrected $50\%$ isocontour volume-of-interest (VOI) is drawn \cite{Koopman2016,Kaalep2018} in order to compute the mean activity concentration $A_{mean}^{(i)}$ in each sphere. The peak activity concentration $A_{peak}$ is computed by placing a spherical VOI with \SI{12}{\mm} diameter such that it yields the hightest mean activity concentration. The recovery coefficients for each sphere $i$ are therefore defined as 

\begin{equation}
    \rmax^{(i)} \; \doteq \; \frac{A_{max}^{(i)}}{A_s} \;, \qquad \rmean^{(i)} \; \doteq \; \frac{A_{mean}^{(i)}}{A_s} \;, \qquad \rpeak^{(i)} \; \doteq \; \frac{A_{peak}^{(i)}}{A_s} \;,  
\end{equation}
where $A_s$ is the \tb\ activity concentration reported in Tab.~\ref{tab:activities}. 

For the coefficient of variation (COV), we define three quadratic region-of-interest (ROI) in three distinct slices in the phantom's background compartment according to Ref.~\cite{Koopman2016,Kaalep2018}. \revised{For each ROI $\mathrm{COV}_j$ $j=1,...9$ is computed by taking the standard deviation of all pixel values and dividing it by the mean $\bar{A}_{j}$.} The COV is the average across all nine $\mathrm{COV}_i$

\begin{equation}
    \mathrm{COV} \; \doteq \; \frac{1}{9} \sum_{j=1}^9 \mathrm{COV}_j \eq \frac{1}{9} \sum_{i=j}^9 \frac{\sigma_{j}}{\bar{A}_{j}} \cdot 100 \;. 
\end{equation}
This is the inverse of the signal-to-noise ratio. $\mathrm{COV}\leq 15\%$ is compliant with the EARL limit for image noise. 

Finally, we consider the average residual lung error by dividing the mean activity concentration in a cylindrical VOI (\SI{120}{\mm} length, \SI{30}{\mm} diameter) placed at the center of the lung insert of the phantom by the average activity concentration $A_{bkg} = 1/9 \sum \bar{A}_{j} $ in the background ROIs.



\section{Results}

In Tab.~\ref{tab:fwhm} we present the FWHM as well as the FWTM form the capillary measurements. 

\begin{table}
    \centering
    \caption{FWHM and FWTM from the capillary measurements in three transverse locations and at the center and $1/8$ of the axial FOV. }
    \label{tab:fwhm}
    {\footnotesize
    \begin{tabular}{llcccccc}
        \toprule 
        Position [\SI{}{\mm}] & Profile & \multicolumn{3}{c}{FWHM [\SI{}{\mm}] } & \multicolumn{3}{c}{FWTM [\SI{}{\mm}] } \\
         & & $1/2$ &  $1/8$   & Average  & $1/2$ &  $1/8$   & Average  \\        \midrule 
        \rowcolor{lightgray} & Radial & 5.95 & 5.95 & 5.95 & 10.81 & 9.93 &	10.37 	  \\
        \rowcolor{lightgray} (0,10) & Tangential & 4.23 & 4.53 & 4.38 & 6.75 & 	6.4 &	6.61 \\
        \rowcolor{lightgray}   & Axial & 7.68 & 5.90 & 6.79 & 18.57 & 12.91  & 15.75  \\
        & Radial & 4.40 & 4.68 & 4.54 & 8.25 & 7.98& 8.11 \\
        (0,100) & Tangential & 	4.48 & 	4.29 & 4.38 & 7.32 & 7.24 &7.28 \\
         & Axial & 6.38 & 6.26 & 6.32 & 21.77 &	11.16 &	16.46 \\
        \rowcolor{lightgray} & Radial & 4.22 & 4.42 & 4.32 & 6.93 &	6.54 & 6.74 \\
        \rowcolor{lightgray} (0, 200) & Tangential & 4.94 & 4.59 & 4.76 & 8.64 & 8.12& 8.38 \\
        \rowcolor{lightgray} & Axial & 8.09 & 7.21 & 7.65 & 24.17 & 11.49& 17.83 \\
        \bottomrule 
    \end{tabular}
    }
\end{table}

\begin{figure}
    \centering
    \includegraphics[width=\linewidth]{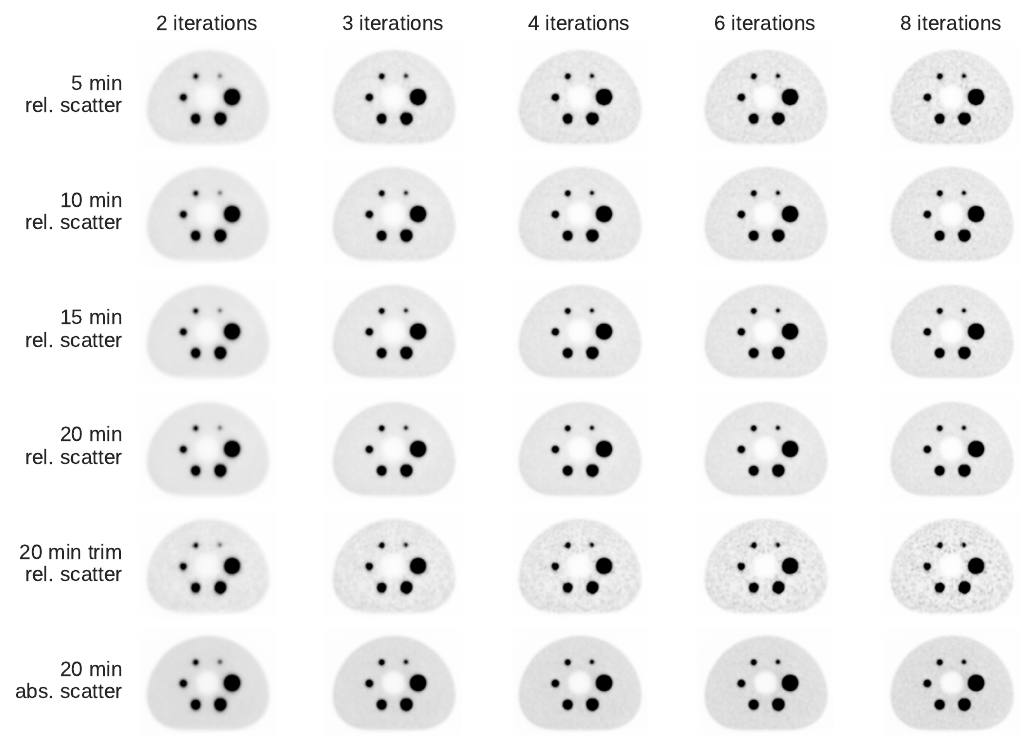}
    \caption{Central slices of the phantom image for different iterations, scan times and scatter correction. }
    \label{fig:slices}
\end{figure}

Fig.~\ref{fig:slices} visualizes the central slice of the NEMA phantom for different iterations, scan times and scatter correction algorithms. At a qualitative level, it is already visible how the noise increases with an increasing number of iterations and with decreasing scan times. The trimmed data is gives the noisiest images. Note that in all cases the \SI{10}{\mm} sphere is clearly visible. 

\begin{figure}
    \centering
    \includegraphics[width=\linewidth]{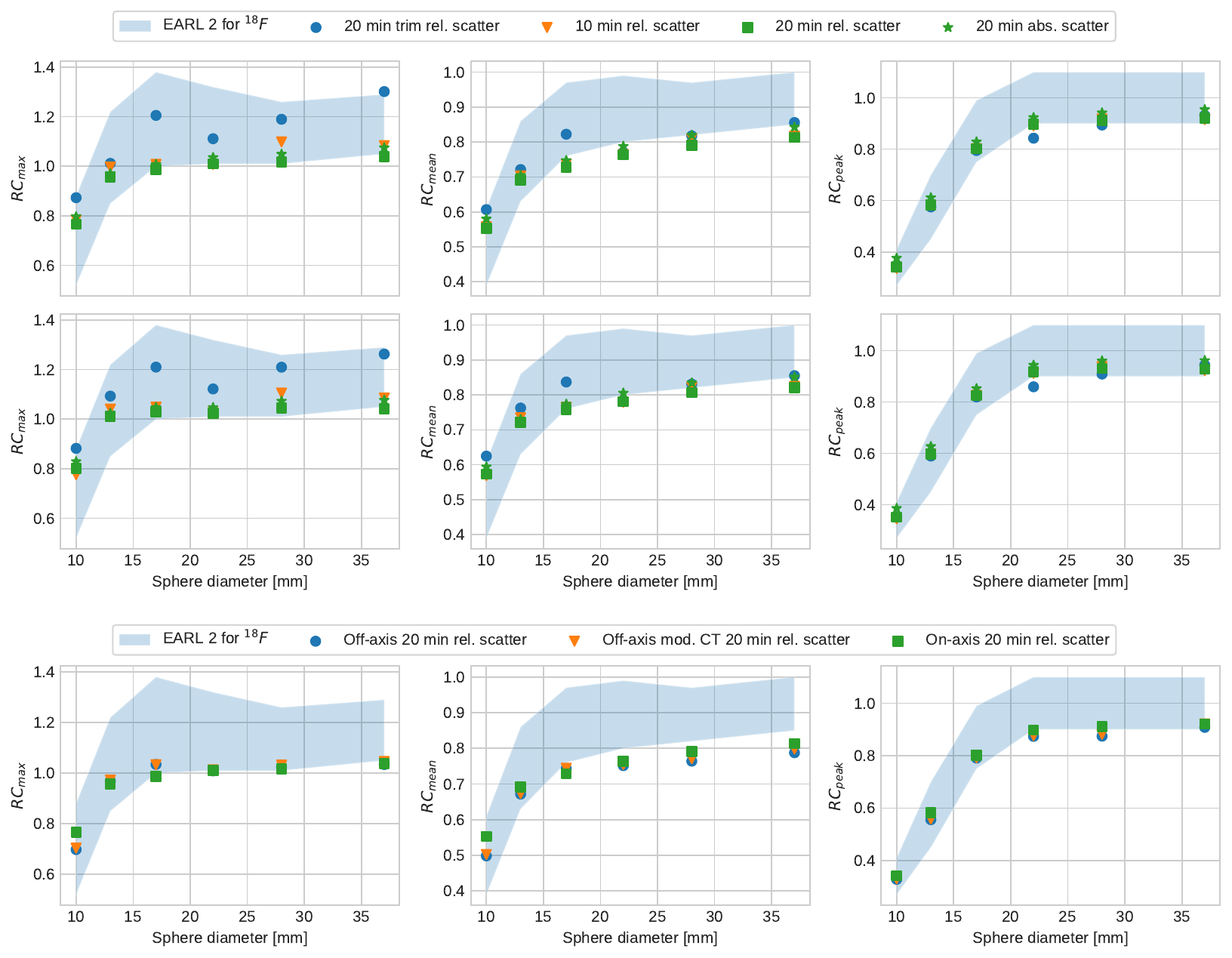}
    \caption{\tb\ recovery for 3 iterations and different scan times and scatter corrections in comparison with the EARL standard 2 for \isotope[18]{F}. \emph{Top row:} recovery coefficients with standard PSF correction  and PR-specific PSF correction. \emph{Center row:} \emph{Top row:} recovery coefficients with PR-specific PSF correction. \emph{Bottom row:} recovery coefficients for on/off-axis phantom and with a corrected AC CT.}
    \label{fig:recovery}
\end{figure}

In Fig.~\ref{fig:recovery} and Tab.~\ref{tab:recovery} we report the recovery coefficients for three iterations in comparison with the EARL standard 2 for \isotope[18]{F}. In the first two rows of Fig.~\ref{fig:recovery}, the three recovery coefficients $\rmax$, $\rmean$ and $\rpeak$ are computed for 10 and \SI{20}{\min} as well as for the \SI{20}{\min} trimmed images (only $10\%$ of the raw data is retained). The dotted line shows the recovery coefficients in the case of absolute scatter correction. In the bottom row, we compare the three \SI{20}{\min} scan with three iterations and relative scatter. The only difference is the location of the phantom and the AC CT: either the phantom is at center of the patient table (\revised{panel (a) of Fig.~\ref{fig:ac_ct}}), at the edge of the patient table (\revised{panel (b) of Fig.~\ref{fig:ac_ct}}) or at the edge of the patient table but with a corrected AC CT (\revised{panel (c) of Fig.~\ref{fig:ac_ct}}). 

\begin{table}
    \centering
    \caption{Recovery coefficients for three iterations. }
    \label{tab:recovery}
    {\footnotesize
    \begin{tabular}{llcccccccc}
        \toprule
        Recovery & Reconstruction & \multicolumn{6}{c}{Single sphere recovery}  & COV [\%] & Lung err. [\%] \\
        & & $\emptyset10$ & $\emptyset13$ & $\emptyset17$ & $\emptyset22$ & $\emptyset25$ & $\emptyset37$ &  &  \\
        \midrule 
        \rowcolor{lightgray}   & \SI{20}{\min} rel. & 0.77 & 0.96 & 0.99 &	1.01 & 1.02 & 1.04 & 12.60 & 6.40 \\ 
        \rowcolor{lightgray}    & \SI{10}{\min} rel. & 0.78 & 1.00 & 1.01 & 1.01 & 1.10 & 1.08 & 14.66 & 6.45 \\ 
        \rowcolor{lightgray}   & Off-axis \SI{20}{\min} rel. & 0.70 & 0.97 & 1.03 & 1.01 & 1.03 & 1.03 & 16.70 	& 8.41 \\ 
        \rowcolor{lightgray}   & \SI{20}{\min} trim rel. & 0.87 & 1.01 & 1.21 & 1.11 & 1.19 & 1.30 & 27.27 & 6.69 \\ 
        \rowcolor{lightgray}  \multirow{-5}*{$\rmax$}  & \SI{20}{\min} abs. & 0.80 & 0.97 & 1.01 & 	1.04 & 1.05 & 1.07 	& 11.88 &  	17.35 \\ 
         & \SI{20}{\min} rel. & 0.55 & 0.69 & 0.73 & 0.76 & 0.79 & 0.81 	& 12.60 	& 6.40 \\ 
         & \SI{10}{\min} rel. & 0.56 & 0.70 	& 0.74 & 0.76 & 0.81 & 0.82 & 14.66 & 6.45 \\ 
          & Off-axis \SI{20}{\min} rel. & 0.50 & 0.67 & 0.74 & 0.75 & 0.76 & 0.79 & 16.70 & 8.41 \\ 
         & \SI{20}{\min} trim rel. & 0.61 & 0.72 & 0.82 & 0.77 & 0.82 & 0.86 & 27.27 & 6.69 \\ 
        \multirow{-5}*{$\rmean$}  & \SI{20}{\min} abs. & 0.58 & 0.70 & 0.75 & 0.79 	& 0.82 	& 0.84 & 11.88 & 17.35 \\ 
        \rowcolor{lightgray}   & \SI{20}{\min} rel. & 0.34 	& 0.58 & 0.80 & 0.90 & 0.91 & 0.92 & 12.60 & 6.40 \\ 
        \rowcolor{lightgray}    & \SI{10}{\min} rel. & 0.34 & 0.58 & 0.80 & 0.89 & 0.92 & 0.91 & 14.66 & 6.45 \\ 
        \rowcolor{lightgray}   & Off-axis \SI{20}{\min} rel. & 0.33 & 0.56 & 0.79 & 0.87 & 0.87 & 0.91 & 16.70 & 8.41 \\ 
        \rowcolor{lightgray}   & \SI{20}{\min} trim rel. & 0.34 & 0.58 	& 0.79 & 0.84 & 0.89 & 0.93 & 27.27 & 6.69 \\ 
        \rowcolor{lightgray}  \multirow{-5}*{$\rmax$}  & \SI{20}{\min} abs. & 0.38 & 0.61 & 0.83 & 0.92 & 0.94 	& 0.95 & 11.88 & 	17.35 \\ 
        \bottomrule
    \end{tabular}
    }
\end{table}

\begin{figure}
    \centering
    \includegraphics[width=\linewidth]{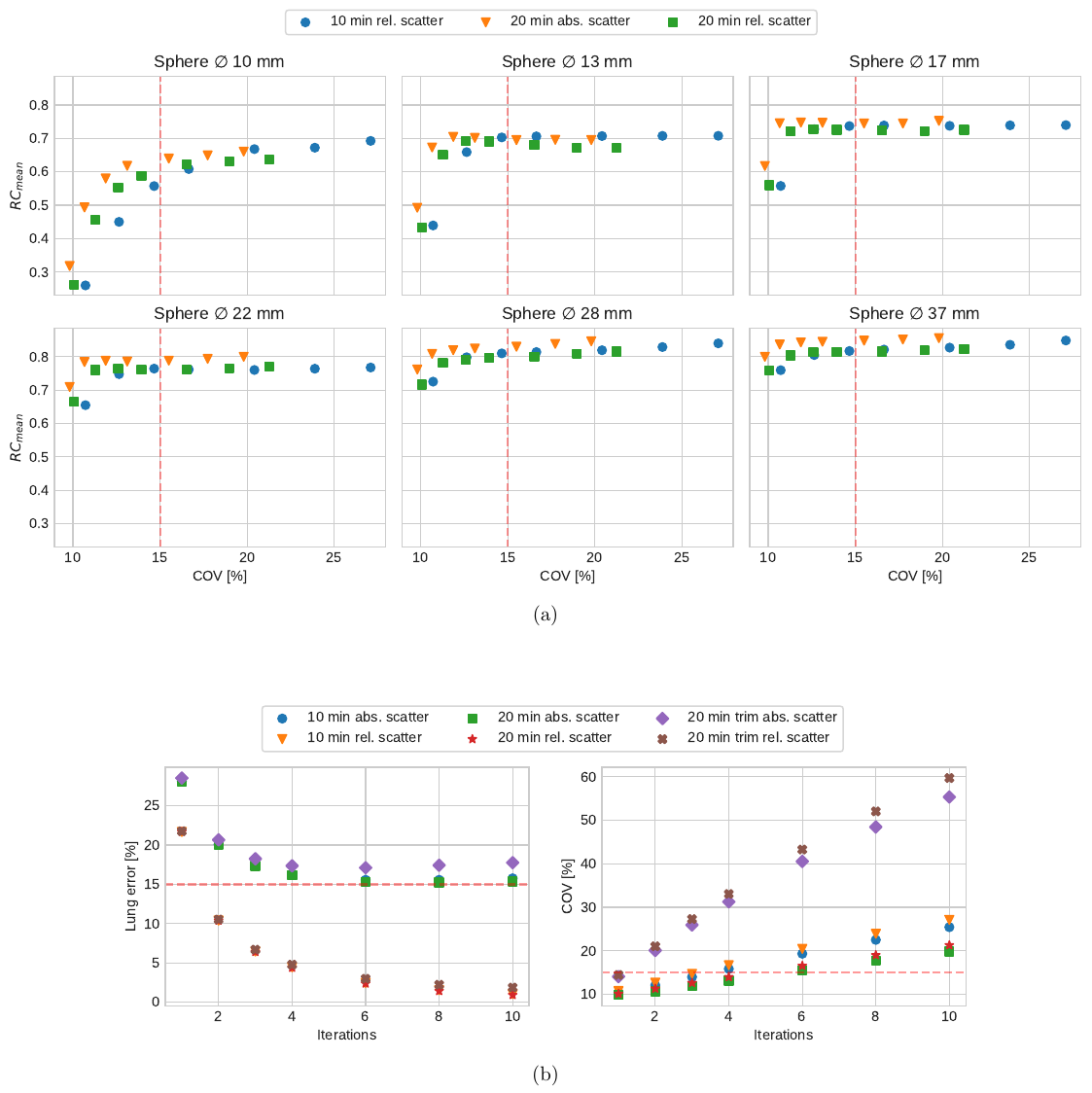}
    \caption{(a) convergence of $\rmean$ for increasing number of iterations as a function $\cov$ for different scan times and scatter correction. (b) lung error and $\cov$ as a function of the number of iterations for different scan times and scatter correction. }
    \label{fig:recovery_convergence}
\end{figure}

\revised{Panel (a) of Fig.~\ref{fig:recovery_convergence} illustrates the convergence of the reconstruction algorithm with increasing number of iterations.} I.e.\ the $\cov$ is plotted against the recovery coefficient $\rmean$ for each sphere of the NEMA phantom. The accepted limit of $15\%$ is marked with the dashed line. The bottom row shows the lung error and $\cov$ as a function of the number of iterations. 

\begin{figure}
    \centering
    \includegraphics[width=\linewidth]{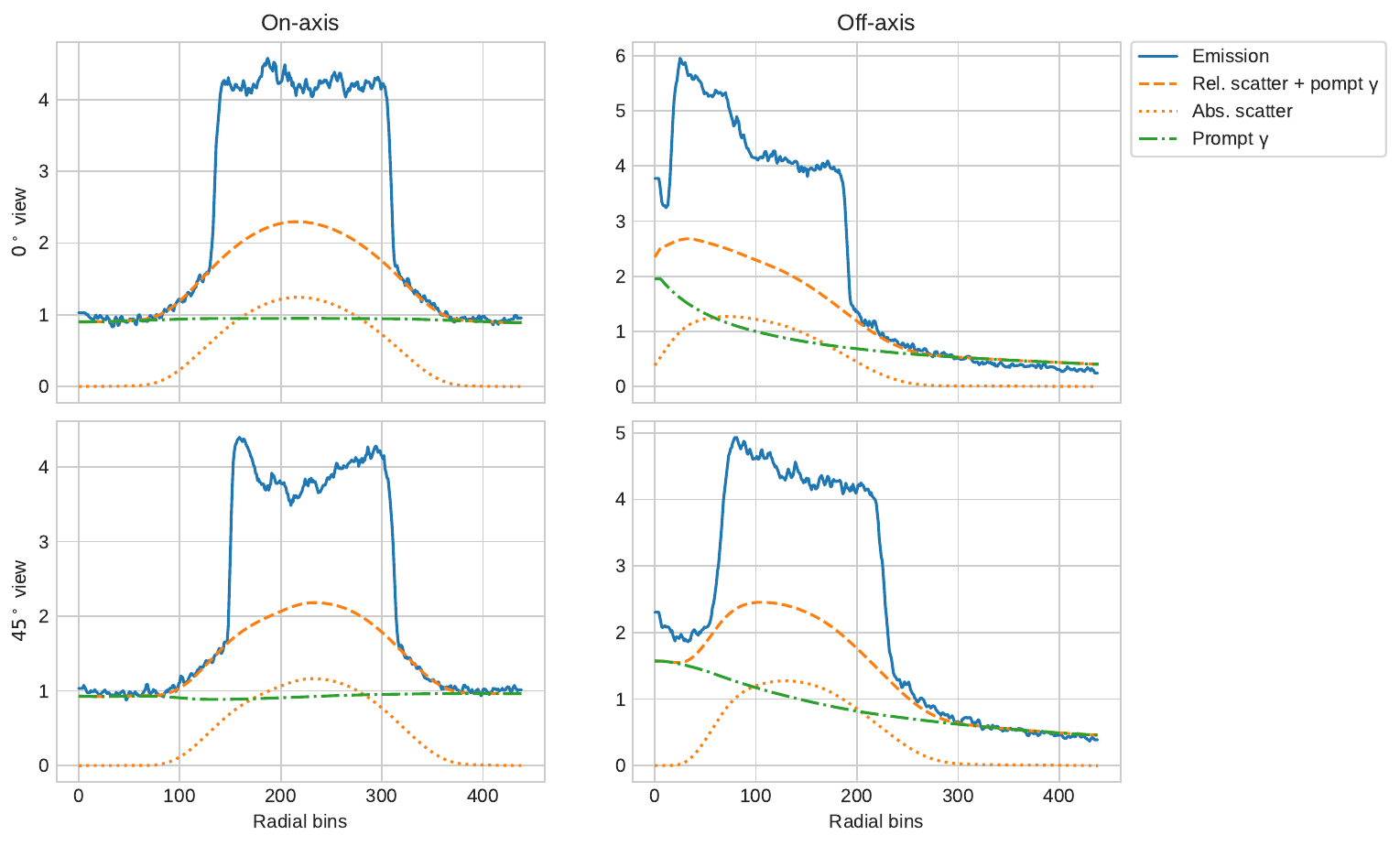}
    \caption{\revised{\emph{Top:} Emission sinogram together with total correction, \pg\ correction and absolute scatter sinograms for the on- (\emph{left}) and off-axis (\emph{right}) measurements at $0^\circ$ view angle (arbitrary unit on y axis). \emph{Bottom:} Same sinograms but at $45^\circ$ view angle.}}
    \label{fig:sinograms}
\end{figure}

Finally, we illustrate on the \pg\ contribution and challenge its correction model. Fig.~\ref{fig:sinograms} illustrates the sinograms at two different view angles (top row is at $0^\circ$ and the bottom row at $45^\circ$ viewing angle). The panels in the left column show the total emission sinogram for the phantom on the central axis of the scanner. \revised{The \pg\ correction (dash-dotted line) dominates the total correction (dashed line) in the tail regions, i.e.\ outside of the phantom.} This highlights the necessity of a correction method for \pg\ when imaging \tb. \revised{The absolute scatter correction (dotted line) falls short in accounting for the correcting the emission sinogram.} \revised{The differences between distorted and the synthetic AC CT for the \pg\ and relative scatter correction from the are below 1\%, i.e.\ could not be visualized in the sinograms of Fig.~\ref{fig:sinograms}.} 

\section{Discussion}

Firstly, we wish to emphasize the transformative potential of imaging an $\alpha$-emitting radionuclide for therapeutic applications with PET resulting in a high qualitative image quality shown in Fig.~\ref{fig:slices}. While imaging of $\alpha$ emitters like \isotope[223]{Ra} or \isotope[225]{Ac} is hampered by the challenging decay radiation (low photon energies and/or BRs) and the limited quantitative abilities of SPECT \cite{Owaki2017,Seo2019,Tulik2024}, \tb\ promises to bring all the advantages of PET to therapies $\alpha$. Our study shows that therapy control and dosimetry of small lesions for $\alpha$ therapies are not wishful thinking anymore, as the images Fig.~\ref{fig:slices} illustrate. 

\revised{\tb\ has a single $\alpha$ emission in its entire decay chain. In the absence of a dosimetry study for humans, we may speculate that the administered activity would have to be few hundreds of \SI{}{\mega\becquerel} in order to achieve a therapeutic effect, in analogy to \isotope[211]{At} \cite{Watabe2025,Wu2026}. The \SI{45.29 \pm 0.10}{\mega\becquerel} total activity in our phantom are therefore a reasonable choice. The simulated low-dose images simulate the case of administering activities like for \isotope[223]{Ra} or \isotope[225]{Ac}, which however have multiple $\alpha$ emission and longer half-lives compared to \tb.}

Fig.~\ref{fig:slices} visualizes how even small lesions, i.e.\ the smallest sphere with $\emptyset \, \SI{10}{\mm}$, can be imaged with a patient-friendly scan time of \SI{20}{\min} and the trimmed data. This can be attributed to the high sensitivity of the Quadra. \revised{In comparison to the phantom study with \isotope[90]{Y}, i.e.\ the only therapeutic radionuclide imaged with PET/CT in clinical routine \cite{Zeimpekis2023}, even at the qualitative level, the image quality obtained with \tb\ exceeds that of \isotope[90]{Y}, despite the longer scan time and higher activity used in the latter. This} is due to the much smaller positron BR of \isotope[90]{Y}. If we dare to compare with \isotope[177]{Lu} imaging, even high-sensitivity ring shaped CZT SPECT/CT struggle to visualize the smallest sphere of the NEMA phantom \cite{Zorz2024}. \revised{It is therefore noteworthy that we can compare the image quality and quantitative accuracy of \tb\ in conjunction with a LAFOV PET/CT with \isotope[18]{F} on SAFOV PET/CT systems. }

Fig.~\ref{fig:recovery} and Tab.~\ref{tab:recovery} demonstrate overall high quantitative accuracy. \revised{Although $\rmean$ and $\rmax$ fall at the lower end of or slightly below the EARL standard 2 limits, this level of quantitative accuracy is particularly encouraging given that the results were obtained from the trimmed raw data.} The HPGe activity measurement can be assumed to be very accurate \cite{Debertin1988}. Possibly, the larger uncertainty of a dose calibrator measurement would increase the overlap of the \tb\ recoveries with the EARL limits.  
\revised{The increased $\cov$ of the trimmed raw data impacts mostly $\rmax$, while leaving only minor differences in $\rmean$ and $\rpeak$.} As expected for a LAFOV PET/CT \cite{Spencer2021,Prenosil2022,Prenosil2022a}, the recover coefficients for the smaller spheres show a relatively higher accuracy, i.e.\ the curves in Fig.~\ref{fig:recovery} are somewhat flatter than the EARL limits. \revised{This highlights again the prospect of imaging small lesions while enabling accurate dosimetry in a future clinical implementation. }

Comparing the top and center rows of Fig.~\ref{fig:recovery}, it is clear that the impact of the post-reconstruction PR correction on the recovery coefficients is marginal. \revised{The maximal improvement on $\rmean$ is 4\% at 3 iterations.} In clinical routine, there is typically no isotope-specific PR correction for \isotope[68]{Ga}, i.e.\ the \isotope[18]{F} PSF correction is applied. \tb\ has a similar PR in water as \isotope[68]{Ga} (see panel (a) of Fig.~\ref{fig:hpge_spectrum}), hence one could also simply neglect the relatively large PR. However, Ref.~\cite{Kertesz2022} showed a significant improvement in  quantitative accuracy for \isotope[124]{I}, which also as a similar PR as \tb\ and \isotope[68]{Ga}, with a dedicated PR correction. Likely, we are unable to see a similar improvement because our \tb-specific PR correction operates only in image-space, which represents a gross simplification compared to the custom projector used in Ref.~\cite{Kertesz2022}'s OSEM implementation. Despite having a PR well shorter than \isotope[82]{Rb} \cite{Lassen2025}, it would be interesting to investigate how much a custom OSEM implementation with \tb-specific projectors would improve the recovery coefficients, in particular given the low statistics environment of our \tb\ measurements compared to the study of Ref.~\cite{Kertesz2022}. We defer exploration of this aspect to future work.

Given the fast convergence of the recovery coefficients and the rapidly increasing $\cov$ with the number of iterations, we deem three iterations as the optimal value. This is merely one iteration less than the typical protocol used in clinical routine for \isotope[18]{F} and accounts for the limited count statistics in our data set.

Despite the quantitative accuracy discussed in the previous paragraphs, there are two main challenges with \tb. Quadra's sensitivity cannot fully compensate the relatively low positron BR of \tb, leading to limited count statistics. The noise quickly rises above the $15\,\% $ threshold for $\cov$ and the lung error even for few iterations, as seen in Fig.~\ref{fig:recovery_convergence}. Nevertheless, our study shows that with a LAFOV PET/CT even activities as low as \SI{4.5}{\mega\becquerel} of \tb\ can be imaged with a reasonable quantitative accuracy. The low positron BR of \tb\ inhibits a similar image quality as presented in Ref.~\cite{Sari2025} for a few \SI{}{\mega\becquerel} of \isotope[18]{F}. \rrevised{If, though unlikely given the single $\alpha$ decay in the whole decay chain, in a future clinical implementation of \tb\ the administered activity} should be as low as for \isotope[223]{Ra} or \isotope[225]{Ac}, we suggest to extend the scan time beyond \SI{20}{\min}.
\revised{We anticipate that scan times on a SAFOV PET/CT system would need to be longer and thus affecting patient management significantly.}
In terms of clinically used radionuclides that face similar challenges for count statistics, \tb\ can probably be best compared to \isotope[124]{I} (see positron BR in Tab.~\ref{tab:isotope}). \revised{However, considering the very long scan time on a SAFOV PET/CT system of Ref.~\cite{Kertesz2022}, it is expected that the increased sensitivity of the LAFOV PET/CT can efficiently compensate for the slightly lower positron BR of \tb.}

The second challenge is the large number of \pg, in particular above the \SI{511}{\kilo\electronvolt} energy window. Considering the left column in Fig.~\ref{fig:sinograms}, the substantial number of observed events observed in the sinograms' tail region, i.e.\ outside the NEMA phantom, demonstrates the high prevalence of random events generated by \pg. However, the linear combination of the (relative) scatter contribution and the smoothed randoms \cite{Conti2016} (dashed line in Fig.~\ref{fig:sinograms}) is an adequate correction model for \pg\ and scatter events in the case of a centered phantom. \revised{Specific physical processes are not separately modeled. False coincidences that are localized outside the phantom are mostly captured by the \pg-scatter model. We can imagine a false coincidence stemming from pair production of a high-energy \pg\ being localized inside the phantom. Such events would not follow scatter kinematics and therefore could lead to a deterioration of recovery coefficients and increased noise. For the centered phantom this, however, does not seem to be a major problem. }
In the case of an off-axis phantom, the limits of this correction model become evident. In particular in the bottom right panel of Fig.~\ref{fig:sinograms}, the dash-dotted line cannot account for full the emission outside of the phantom on the detector-proximal side. The rise in the emission sinogram towards the detector ring cannot be represented by uncorrelated events, i.e.\ smoothed randoms, nor scatter kinetmatics. Possibly, also increased pile-up of \pg\ below the \SI{511}{\kilo\electronvolt} energy window leads to an underestimation of the total correction of the emission sinogram.  

\revised{The absolute scatter correction algorithm fails to account for \pg\ contributions in the emission sinogram (see Fig.~\ref{fig:sinograms}).} Neither the decreasing emission tails close to the phantom walls or the flat emission background closer to the detector rings are appropriately modeled. In terms of quantification, absolute scatter correction seems to be slightly better as a significantly lower number of events is subtracted from the emission sinogram. However, given the large number events outside the source region shown in the tails of Fig.~\ref{fig:sinograms} we strongly discourage using an absolute scatter correction for \tb. 

Interestingly, the distorted AC CT does not significantly affect the quantitative accuracy (see bottom row of Fig.~\ref{fig:recovery}). In the right column of Fig.~\ref{fig:sinograms}, the difference between the sinograms with distorted and the modified CT sinograms is marginal. Likely, this is due to the fact that the mask used for the fitting of the relative scatter and \pg\ contribution lies entirely on the detector-distal side of the phantom. 

\revised{The clinical translation of \tb\ remains contingent on several prerequisites beyond imaging performance. Firstly, production has to be scaled \cite{favaretto2024,Tosato2025} in order to allow for extensive preclinical studies beyond Refs.~\cite{Umbricht2019,Mapanao2025} and eventually clinical trials. As seen with e.g.\ \isotope[211]{At} or \isotope[223]{Ra}, however, this will require substantial resources for in terms of infrastructure and logistics. Furthermore, the specific details of a patient protocol will have to be defined within the framework of a clinical trial. We do not anticipate \tb\ to replace other therapeutic radionuclides, but rather as an additional tool for therapeutic applications in nuclear medicine, where the treatment and imaging of small lesions is required. 
Given the short half-life compared to other therapeutic radionuclides, we envision an outpatient regime for therapies with \tb. Depending on the tracer, uptake is rather fast and post-treatment imaging (even multiple time points for dosimetric purposes) could be performed within few hours post-injection. However, the national radiation protection authorities will have to specify the constraints for a safe handling, administration to the patient and waste management. }

\revised{Looking beyond standard PET imaging, \tb\ can be used for positronium lifetime imaging as shown in Ref.~\cite{Mercolli2026b}. Given the large of \pg\ emission, also SPECT and Compton imaging may be imaginable. }\rrevised{However, the high \pg\ energies will pose challenges to modalities that rely on collimation.  }


\section{Conclusions}

In this study we demonstrated that \tb\ can be imaged with a commercial LAFOV PET/CT. Despite the low positron BR and a high number prompt $\gamma$, it is possible to image \tb\ with a quantitative accuracy that is comparable to \isotope[18]{F}. Accurate dosimetry and post-treatment verification should be feasible even for small lesions in a future clinical translation of \tb. Our findings strengthen the case of \tb\ as a promising radionuclide for targeted $\alpha$ therapy and may open new avenues for its application in theranostics.

\section*{Acknowledgments}

The authors thank Hasan Sari, William M. Steinberger and Mohammadreza Teimoorisichani for useful discussions and for their help with e7tools. They also acknowledge the help of Ângelo Rafael Felgosa Cardoso, Angela Filipa Silva Mendes and Marco Viscione for the scheduling of the experiments at Inselspital. The authors thank Stuart Warren, Maryam Mostamand, Reinhard Heinke, Wiktoria Wojtaczka and ISOLDE (experiment IS688), CERN, in particular the target group,  RILIS group, health physics and CERN/PSI logistics. Last, but not least, they are grateful for support from the PSI radiation safety group.

\section*{Declarations}

\subsection*{Funding}
\noindent This research is partially supported by the grants no. 216944 and no. 224901 of the Swiss National Science Foundation.

\subsection*{Competing interests}
\noindent All authors have no conflict of interests to report.

\subsection*{Ethics approval}
\noindent Not applicable. 

\subsection*{Consent to participate}
\noindent Not applicable. 

\subsection*{Data availability}
\noindent The datasets generated and/or analyzed in the current study are available from the corresponding author upon reasonable request.

\subsection*{Authors' contributions}
\textbf{LM}: Conceptualization, Formal analysis, Investigation, Methodology, Software, Validation, Writing - Original Draft, Writing - Review \& Editing
\textbf{PG and AM:} Investigation, Resources, Writing - Review \& Editing
\textbf{SB:} Project administration, Investigation, Supervision, Writing - Review \& Editing
\textbf{LE:} Investigation, Writing - Review \& Editing
\textbf{UK:} Resources, Writing - Review \& Editing
\textbf{NvdM:} Conceptualization, Investigation, Supervision, Project administration, Writing - Review \& Editing

\bibliographystyle{elsarticle-num} 
\biboptions{sort&compress} 
\bibliography{tb149_bibliography.bib}



\end{document}